\documentclass[runningheads]{llncs}
\usepackage[T1]{fontenc}
\usepackage[misc]{ifsym}
\usepackage{graphicx}
\usepackage{amsmath}
\usepackage{amsfonts}
\usepackage{subcaption}
\usepackage{adjustbox}
\usepackage{makecell}
\usepackage{hyperref}
\usepackage{color}

\begin{document}
\title{Matheuristic Local Search for the Placement of Analog Integrated Circuits}
%
%

\author{Josef Grus\inst{1,2(\textrm{\Letter})}\orcidID{0000-0002-1136-370X} \and Zdeněk Hanzálek\inst{2}\orcidID{0000-0002-8135-1296}}

%
\authorrunning{J. Grus and Z. Hanzálek}
%
\institute{DCE, FEE, Czech Technical University in Prague, Czech Republic \email{grusjose@fel.cvut.cz} \and
IID, CIIRC, Czech Technical University in Prague, Czech Republic
\email{zdenek.hanzalek@cvut.cz}}

\maketitle              
\begin{abstract}
The suboptimal physical design of the integrated circuits may not only increase the manufacturing costs due to the larger size of the chip but can also impact its performance by placing interconnected rectangular devices too far from each other. In the domain of Analog and Mixed-Signal Integrated Circuits (AMS ICs), placement automation is lacking behind its digital counterpart, mainly due to the variety of components and complex constraints the placement needs to satisfy. Integer Linear Programming (ILP) is a suitable approach to modeling the placement problem for AMS ICs. However, not even state-of-the-art solvers can create high-quality placements for large problem instances. In this paper, we study how to improve the results of our previous ILP model, first by introducing additional constraints and second by using matheuristics. Given the initial solution we obtain using our original ILP model, we use the solver to perform a local search. We try to improve the criterion by considering only a few spatially close rectangles while keeping the rest of the placement fixed. This local search approach enables us to significantly improve the quality of instances whose solution space we could not sufficiently explore before, even when the computation time reserved for the matheuristic is limited. Finally, we evaluate our revised approach on synthetically generated instances containing more than 200 independent rectangles and on real-life problems.
\end{abstract}
\section{Introduction}
The importance of ICs for modern civilization is apparent. Advanced computing, Internet-of-Things devices, automotive, and consumer electronics rely on high-performance ICs. Such market pressure further motivates the companies to shorten the design time and lower the development costs to increase their profitability and strengthen their market position. One of the crucial steps in the design of the ICs is the physical design. During this step, the circuit diagram is converted into the geometrical representation of the final product - positions and orientations of the rectangular devices (transistors, resistors, etc.) are determined during the placement phase, and the interconnections between them are planned during the routing phase. While these two steps are commonly solved one after another, the placement phase needs to consider the approximated interconnections to make the final product competitive and high-performant.

AMS components remain crucial nowadays, as operational amplifiers and analog-to-digital converters are required to convert signals from many sensors surrounding us. The placement phase for the digital ICs has already been successfully automated. Digital devices are in the form of standardized cells, each sharing the same height, and they are placed in rows rather than freely. These properties make the digital ICs' placement similar to the 1D bin packing problem and enable the automation tools to handle thousands of devices. 

On the other hand, AMS ICs usually contain tens or hundreds of devices. However, the devices may appear in different sizes and aspect ratios and can be placed freely. They also have different voltage levels, which does not happen in a digital domain. Furthermore, the presence of noise and other negative effects inherent to the analog domain significantly influence the overall performance of the circuits. This is mitigated by additional constraints and rules the engineers must adhere to. Due to these complications, the placement of the AMS ICs has not been largely automated and still remains a time-consuming and error-prone manual process; its automation is pursued by projects funded both by DARPA \cite{darpa} and EU \cite{ambeation}. It is further complicated by constraints specific to different technologies of the ICs. This paper specifically discusses BCD technology (technology combining analog, digital, and high-voltage components), which means the placer has to consider various minimum distances between devices and isolated pockets, among other features.

ILP offers a formalism to successfully model the placement problem of AMS ICs. Most constraints regarding the sizes of the devices and their mutual proximity or connectivity can be described using linear inequalities, while the non-linear criterion of the circuit's area might be approximated with its half-perimeter. Nevertheless, even the state-of-the-art ILP solvers, which improve every year, cannot sufficiently well explore the space of feasible placements of larger ICs. 

In this paper, we build upon our previous work \cite{icores23}, where we used warm-started ILP to place devices of the AMS ICs. We discuss the effect of additional symmetry-breaking and redundant constraints on the model's performance. Finally, we develop a Matheuristic (MH) local search technique, which iteratively optimizes the initial solution obtained by solving the entire model, and which offers significant improvement, especially on the large synthetically generated instances with more than 200 devices to be placed. This paper is structured as follows. In Section~\ref{related_work}, we mention the previous work done in domains of both placement and matheuristics. In Section~\ref{problem_formulation}, we formulate the placement problem for BCD technology. Section~\ref{ilp_model_section} describes our original ILP model, as well as additional redundant constraints we experimented with. Section~\ref{matheuristic_section} describes our MH approach. In Section~\ref{experiments}, we describe the problem instances and present the experimental results, which show how well the MH approach performs. Also, real-life instances are evaluated and compared with manual benchmarks. Finally, conclusions are drawn in Section~\ref{conclusion}.

\section{Related Work}\label{related_work}
Even though the placement of the AMS ICs is not as automated as in the case of digital ICs, the problem has already been tackled in the past. Many methods use so-called topological representation - the solution is encoded using relative positions between the devices. Then, a packing procedure is used to convert the representation into the actual placement. Sequence pairs are one such representation. Proposed in \cite{murata_sequence_pairs}, two permutations of the devices encode the relative positions between devices. Importantly, as was demonstrated in \cite{handling}
, this formulation can be extended to successfully model symmetry groups and other crucial features. Another example of the topological representation is B*-trees, which use binary trees to determine the relative positions between the parent and child nodes. Used in \cite{ga_with_trees,representation_as_trees}, this representation offers a low level of redundancy in its search space.

Other methods consider the absolute coordinates of the devices. This makes encoding constraints such as symmetry groups easier; however, it also introduces infeasible solutions to search space. In the early work of \cite{koan_anagram}, the simulated annealing was used to optimize the coordinates of the devices. The criterion contained both the area and wire length of the IC, as well as penalty terms for constraint violations. In \cite{martins_absolute}, a similar approach, using a multi-objective constrained variant of simulated annealing, was also considered. Alternatively, methods described in papers \cite{ntuplace,eplace} firstly use the global placement phase, where the approximate positions of the devices are determined using non-linear programming
, and then the feasible placement without the overlaps is created using Linear Programming (LP). The mentioned core was extended to accommodate the different manufacturing layers of the ICs in \cite{gaafteropt}. In \cite{date_placer_fdgd}, the neural network was used to estimate the circuit's performance, and it was added to the differentiable criterion. 

The force-directed approach was successfully applied to placement in \cite{kraftwerk_fdgd}, where the attractive and repulsive forces between the devices were derived from the connectivity of the IC and the devices' overlaps, respectively. Machine learning found its applications as well. An end-to-end pipeline of \cite{graph_placement} was utilized as a placer of macros, while the learned model performed fine-optimization of the already-placed IC in \cite{date_rl}. 

While the methods outlined in the previous paragraphs successfully solved their associated placement problems, we cannot directly apply them to BCD technology ICs; these ICs rely on various minimum allowed distances between devices, isolated pockets, and other features that were rather omitted in the previous works. This was also a reason why we used the ILP, which allowed us to model these crucial features easily.

The ILP was applied to placement problems in the past. In \cite{ilp_solver}, the authors used hierarchical decomposition to improve the solver's performance and created high-quality placements. In our previous work \cite{icores23}, we employed Force-Directed Graph Drawing-based (FDGD) method to warm start the solver instead of relying on decomposition. Our proposed MH offers to improve the results produced by other methods even when the warm starting the solver or decomposing the problem is not sufficient or leads to low-quality solutions. Furthermore, ILP is often used to solve subproblems that arise within the problem of placement, such as the determination of the number of fingers of transistors \cite{nlp-abs-app-includes-wpe}.

The placement of AMS ICs much resembles other problems encountered within the domain of operations research. Rectangle packing can be viewed as a simplification of this paper's topic due to the rectangular shape of the devices. Papers \cite{CSP_ortho,CSP_abs_rela} used constraint programming to solve the rectangle packing problem. In \cite{Hopper1999AGA}, a genetic algorithm was used together with a Bottom-Left first packing heuristic. Later, the GRASP metaheuristic was applied to strip packing \cite{grasp_packing}. Even more closely related to our problem is Facility Layout Problem (FLP), where the task is to determine the positions of the facilities while minimizing the travel distances between them. This can be perceived as an analogy to the interconnectivity of the devices. ILP formulations of the FLP were investigated in \cite{opt_fac_mip,fac2}. The latter work optimized the paths between the departments simultaneously with the layout, which resembles the simultaneous optimization of placement and routing in the case of ICs. 

MHs, heuristics based on mathematical programming, have been recently successfully applied to many combinatorial problems \cite{Fischetti2016mhchapter}, especially with the ever-increasing performance of the black-box ILP solvers. While the solvers often cannot solve the industrial-size instances, their search capabilities when the model is smaller cannot be ignored. The MHs were used successfully in the domains of scheduling or routing, but the literature regarding their use for packing and cutting is rather sparse \cite{poly2022cuttingMH}. There are many ways how to build the heuristic around the ILP solver. The constructive MHs iteratively solve a series of simpler subproblems and construct the final solution by combining the intermediate results. This was used both for rostering problems \cite{Smet2014mh_const}, as well as for FLP \cite{fac_mip_but_nonexact}. In the latter, authors fix the relative positions between the already placed facilities and iteratively add the remaining ones until the layout is completed. Evolutionary MHs use mathematical programming to tackle the efficiently solvable subproblems encountered while using metaheuristics. In \cite{monch2018mhgabatch}, parallel batch processing scheduling is tackled using a genetic algorithm, and LP is used to improve the solution by solving the minimum cost flow problem. 

Finally, the MHs are often used to perform the local search. Given a starting solution to a problem, we try to improve it by solving the restricted variant of the original ILP model. There are several ways how to achieve such restriction
. The first way, called local branching, limits how many variables can change its value. Assuming the ILP model only contains binary variables, then the following constraint can be introduced \cite{Fischetti2016mhchapter}:
\begin{equation}
    \sum_{i \in B_0}x_i + \sum_{i \in B_1} (1-x_i) \le k
\end{equation} where variable $x_i$ was originally assigned to 0, if $i \in B_0$ and vice versa. The restrictiveness depends on the value of $k$. Local branching was successfully used in the improvement phase of \cite{Smet2014mh_const}. In \cite{Yang2021localbranching}, the flow-shop problem with time windows was tackled, and local branching was even used to construct the feasible solution from the initial infeasible one. 

Another way to restrict the search space is to explicitly fix a subset of variables of the model. This application is very close to Large Neighborhood Search \cite{LNS2021matheur} or Ruin and Recreate heuristics \cite{berghe2020ruinrecreate}; the damaged solution (i.e., the free variables in the restricted ILP model) is repaired using the exact solver. Variable-fixing local search MHs were successfully applied to the scheduling domain, such as in the case of university timetabling \cite{lindalh2018FixAOpt}, flow-shop scheduling \cite{Croce2014mh_varfix_sched}, and evacuations scheduling \cite{Tkindt2016evacuations}. In these works, the choice of free and fixed variables is crucial for the successful application of MHs and often depends on domain-specific information. In this paper, we decided to apply such variable-fixing MH to our placement problem.

\section{Problem Formulation}\label{problem_formulation}
\begin{figure}
    \centering
    \includegraphics[width=0.62\textwidth]{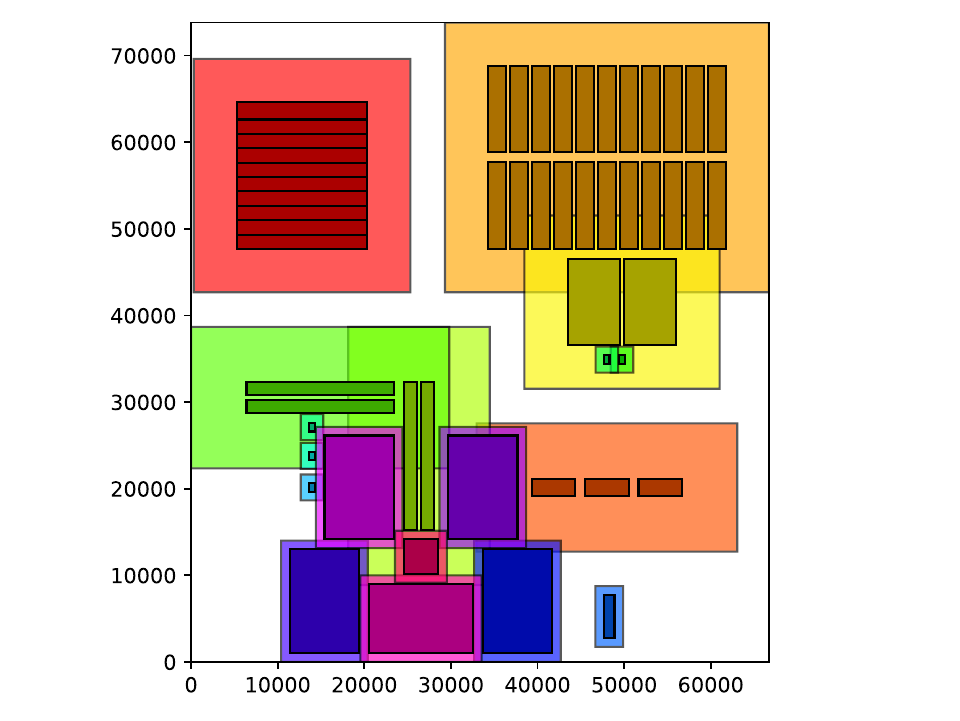}
    \caption{Example placement with critical constraints of the BCD technology \cite{icores23}.} 
    \label{fig:example_placement}
\end{figure}

During the placement phase of the physical design of the AMS ICs, the positions and orientations of the devices are determined. The input of the problem, the netlist, contains information about the sizes of the devices, their voltage level, and interconnectivity. The devices have a rectangular shape of fixed size and can be rotated. Furthermore, we need to consider topological structures. These are higher-level building blocks, such as differential pairs or current mirrors, and they consist of several devices that have to be placed in a regular pattern (see two columns of darker rectangles in Fig.~\ref{fig:example_placement}). Thus, we enumerate all possible variants (with a varying number of rows and columns into which the devices are organized) of such topological structures beforehand, using algorithms based on list scheduling \cite{icores23}. Afterward, we treat both the single devices and the topological structures as rectangles with multiple variants (in the case of single devices, the only alternative variant is rotation). Further in the text, we refer to both types of these building blocks as rectangles. Given a task to place $n$ rectangles, we describe each one of them with the coordinates of its bottom-left corner $(x_i,y_i)$ and its size $(w_i,h_i)$, which corresponds to one of its $m_i$ variants.

Since we want to create as small a placement as possible, we would like to minimize its area $W\cdot H$. However, due to our use of ILP, we minimize the half perimeter of the placement's bounding box $W+H$ instead.

The overall connectivity is modeled as Half Perimeter Wire Length (HPWL). The core concept of connectivity is a set of nets $E$ - each net $e \in E$ consists of a set of connected rectangles $L_e$. Each rectangle can be a member of multiple nets. The overall connectivity is formulated as follows:
\begin{equation}
    \mathrm{HPWL} = \sum_{\forall e \in E} c_e \cdot \left( \max_{i \in L_e} x_i^c -  \min_{i \in L_e} x_i^c + \max_{i \in L_e}y^c_i -  \min_{i \in L_e}y^c_i\right)
\end{equation}
where the centroid coordinates are given by:
\begin{equation}
    x^c_i = x_i + w_i/2
\end{equation}
\begin{equation}
    y^c_i = y_i + h_i/2
\end{equation}

Multiplied by its cost $c_e$, each net contributes to the overall HPWL metric the half of the perimeter of the smallest bounding box that contains all of the net's rectangles' centroids \cite{gaafteropt}. Altogether, our task is to find a feasible placement that not only minimizes the area of its bounding box but minimizes the HPWL metric as well.

The physical devices (darker rectangles surrounded by lighter shells in Fig.~\ref{fig:example_placement}), such as transistors, cannot overlap when they are manufactured in the same layer. Furthermore, an increased minimum distance can be imposed between some devices, e.g., to mitigate the effect of the noise on sensitive components. We also need to model additional empty space, or pocket, around the placed structures and devices (the lighter shells around packed devices in Fig.~\ref{fig:example_placement}). Pockets are needed to isolate devices with different voltage levels, which is common for BCD technology. When the devices do not share their input voltage (BULK) net, and thus their voltage level may differ, we need to place them so their pockets do not overlap. Otherwise, their pockets can be merged as long as their internal devices do not overlap, as the yellow and orange rectangles in Fig.~\ref{fig:example_placement} demonstrate.

Additional constraints include the control of the aspect ratio of the final placement. Also, the engineer can restrict a subset of rectangles from a part of a canvas; we call this type of constraint a blockage area. An example is shown in the bottom-left corner of Fig.~\ref{fig:example_placement}, which remained unoccupied due to the use of the blockage area. Furthermore, a group of rectangles may belong to a symmetry group, which shares a common axis of symmetry. An example is a group of darker rectangles with the vertical axis of symmetry located in the bottom part of Fig.~\ref{fig:example_placement}.

\section{ILP Model and Extensions}\label{ilp_model_section}
\subsection{Baseline Model}\label{baseline_model}
We use our model proposed in \cite{icores23}, which was extended from rectangle packing formulation in \cite{CSP_ortho}. Let $\mathcal{I} = \lbrace 1, \dots, n \rbrace$ be set of rectangles' indices. Four real variables represent each rectangle; coordinates of its bottom-left corner $(x_i,y_i)$ and width and height $(w_i, h_i)$, which has to correspond to one of the $m_i$ pre-defined variants $(w_{i}^{k}, h_i^{k}),~k\in \left\{1,\dots,m_i\right\}$. Note that the sizes of rectangles' variants are increased to model the use of the pockets. The selection of variants is made using binary variables $s_i^k$ for each rectangle $i$ and variant $k$, as is shown in equations \eqref{eq:var0}, \eqref{eq:var1}. $k$-th variant is selected if $s_i^k=1$. Placement's width $W$ and height $H$ are variables constrained by the positions of the placed rectangles.

\begin{align}
	\label{main}
	& x_i + w_i \le W,~~ y_i + h_i \le H&\forall i \in \mathcal{I}\\
	& \sum_{k=1}^{m_i} s_i^{k} = 1& \forall i \in \mathcal{I}\label{eq:var0}\\
	& w_i = \sum_{k=1}^{m_i} w^k_i \cdot s_i^k,~~h_i = \sum_{k=1}^{m_i} h^k_i \cdot s_i^k &\forall i \in \mathcal{I}\label{eq:var1}\\
	& \sum_{k=1}^{4} r^k_{i,j} \ge 1 &\forall i, j \in  \mathcal{I}:~i < j \label{pos0} \\
	& x_i + w_i + a_{i,j}\le x_j + M  (1 - r^1_{i,j}) &\forall i, j \in  \mathcal{I}:~i < j\label{pos1} \\ 
	& y_i + h_i + a_{i,j} \le  y_j + M  (1 - r^2_{i,j}) &\forall i, j \in  \mathcal{I}:~i < j\\
	& x_j + w_j + a_{i,j} \le x_i + M  (1 - r^3_{i,j}) &\forall i, j \in  \mathcal{I}:~i < j\label{pos2} \\ 
	& y_j + h_j + a_{i,j}\le  y_i + M (1 - r^4_{i,j})  &\forall i, j \in  \mathcal{I}:~i < j\label{rel2_eq}\\
	& x_i, y_i, w_i, h_i \ge 0 & \forall i \in  \mathcal{I}\\
	& W,~H \ge 0 \\
	& s^k_i \in \lbrace 0, 1 \rbrace & \forall i \in \mathcal{I}~\forall k \leq m_i \\
	& r^k_{i,j} \in \lbrace 0, 1 \rbrace \notag&\forall i, j \in  \mathcal{I}:~i < j \\
	&& \forall k \in \lbrace 1,2,3,4\rbrace\label{last}
\end{align}

Non-overlapping of the devices is ensured by binary variables $r_{i,j}^k$ and inequalities \eqref{pos0} - \eqref{rel2_eq}, which utilize the big-M approach \cite{bigM}. At least one of the inequalities, which corresponds to the relationship (left/right/over/under) between rectangles, must be valid ($r_{i,j}^k = 1$). Parameter $a_{i,j}$ defines the minimum allowed distance between rectangles. By setting the parameter $a_{i,j}$ to the negative value, the solver can place associated rectangles with their pockets merged, similarly to device layer-aware placements \cite{gaafteropt}. Ultimately, the ILP model for feasible placement of $n$ rectangles uses $\sum_{i=1}^n m_i$ binary variables to encode variant selection, and $4 \cdot \binom{n}{2} = 2 \cdot n \cdot (n-1)$ binary variables to encode the relative positions between rectangles.

Blockage areas are modeled as additional dummy rectangles. We fix their positions and sizes and define the minimum allowed distance parameters. $a_{i,b} = 0$ if the rectangle $i$ is blocked by the blockage area $b$; if the rectangle is unaffected by the blockage area $b$, we simply omit the associated relative position constraints from the model. 

We define the final aspect ratio as $\mathrm{AR} = \min\left\{W,H\right\} / \max\left\{W,H\right\}$, and we want to ensure that $l_R \le AR \le u_R$ holds for chosen aspect ratio parameters $ 0 \le l_R \le u_R \le 1$. Then, the following additional constraints are needed. The binary variable $r_R$ is used to handle the non-convex solution space that is induced when $u_R \neq 1$. When $u_R=1$, we omit the associated inequalities entirely.
\begin{align}
	& l_R \cdot W \le H  \le u_R \cdot W + M \cdot (1-r_R)  \\
	& l_R \cdot H \le W \le u_R \cdot H + M \cdot r_R\\
	& r_R \in \lbrace 0; 1 \rbrace
\end{align}

To model the symmetry groups, we require another continuous variable per group to represent the axis of symmetry. Assume that $G$ is the symmetry group with the vertical axis of symmetry, whose horizontal position is determined by the real variable $x_G$. The symmetry group consists of self-symmetric rectangles $(i,-)$ and symmetric pairs $(i,j)$. Then the following equations constrain the symmetry group's rectangles to share the same axis of symmetry:
\begin{align}
    w_{i} &= w_{j}& \forall (i,j)\in G \\
    h_{i} &= h_{j}& \forall (i,j)\in G \\
    y_{i} &= y_{j}& \forall (i,j)\in G\\
    x_{i} + x_{j} + w_{i} &= 2 \cdot x_G  & \forall (i,j)\in G \\
    2\cdot x_{i} + w_{i} &= 2\cdot x_G & \forall (i, -) \in G 
\end{align}

HPWL connectivity elements are formulated per net. Thanks to the minimization of the connectivity in the final criterion, no integer variables are needed. For each net $e$, we create four continuous variables $X^M_e, X^m_e, Y^M_e, Y^m_e \in \mathbb{R}$, which describe the net's bounding box. Then, we formulate the connectivity criterion $\mathcal{L}_C$ using the following constraints for each net $e \in E$, given the set of the net's connected rectangles $L_e$ and net cost $c_e$:
\begin{align}
    X^M_e \ge x_i + w_i/2 & & \forall i \in L_e \\
    X^m_e \le x_i + w_i/2 & & \forall i \in L_e \\
    Y^M_e \ge y_i + h_i/2 & & \forall i \in L_e \\
    Y^m_e \le y_i + h_i/2 & & \forall i \in L_e 
\end{align}
\begin{equation}
    \mathcal{L}_\mathrm{C} = \sum_{\forall e \in E} c_e \cdot \left( X^M_e - X^m_e + Y^M_e - Y_e^m\right)
\end{equation}

To minimize the area of the placement, which is a non-linear expression $W\cdot H$, we approximate it using the half perimeter of the placement's bounding box: 
\begin{equation}
    \mathcal{L}_A = W + H
\end{equation}

We expect that thanks to the correlation between the perimeter and the area of the bounding rectangle, a solution minimizing $\mathcal{L}_A$ will have a small area as well. Ultimately, the final criterion function is defined as:
\begin{equation}
    \mathcal{L} = c_A \cdot \mathcal{L}_A + \frac{c_C}{\sum_{\forall e \in E}c_e}\cdot\mathcal{L}_\mathrm{C}
\end{equation} where the $c_A$, $c_C$ are tunable costs; by tuning them, we can achieve a suitable trade-off between both $\mathcal{L}_A$ and $\mathcal{L}_\mathrm{C}$. However, since there are only two criterion elements, we fix $c_A=1$ and tune only the connectivity cost. Furthermore, we divide $\mathcal{L}_C$ by $\sum_{\forall e \in E} c_e$, so the effect of using a specific value of $c_C$ is less sensitive to a number of nets present in the IC.

\subsection{Improving the Performance of the Solver}
As we have shown in \cite{icores23}, the presented formulation leads to feasible high-quality placements, but the performance of even the state-of-the-art ILP solvers is insufficient when the number of rectangles grows. We were able to mitigate this problem by providing a solver with an FDGD-based solution as a warm start. In this paper, we want to go even further, and we try to introduce redundant constraints to the original model that do not affect the optimal solutions but could potentially improve the performance of the solver.

\subsubsection{Symmetry Breaking}\label{sb_ilp}
Firstly, we tried to remove the symmetric solutions from the search space. Since all of our constraints are rotation invariant (with the only exception being symmetry groups), we can prune the search space by fixing the orientations or positions of specific rectangles. Firstly, we select a suitable rectangle (the largest one as in \cite{CSP_abs_rela}); let its index be $K$. Then, to remove the solutions symmetrical with respect to the $y=x$ axis, we set all variant variables of rectangle $K$, which correspond to a rotated variant with index $r$, to zero.

The second approach is concerned with the solution symmetry achieved by swapping the quadrants of the bounding box. For example, from the current solution, another one can be created by simply mirroring it with respect to either $x=\frac{W}{2}$ or $y=\frac{H}{2}$ axes, or by reflecting it with respect to point $(\frac{W}{2}; \frac{H}{2})$ point. To prune these parts of the search tree, we constrain the coordinates of rectangle $K$ so its centroid lies within the first quadrant, closest to the origin:
\begin{align}
	2 \cdot x_K + w_K \leq W \\
	2 \cdot y_K + h_K \leq H
\end{align}

\subsubsection{W+H Constraint}\label{wh_ilp}
If we could predict how large the bounding box of the optimal solution would be, we could prune the search space using constraint:
\begin{equation} 
	W + H \leq P
\end{equation} where $P$ is the upper bound on the half perimeter of the solution obtained from the prediction. There are two reasons why this could improve the performance of the model. Firstly, such a hard constraint prunes some branches of the search tree that would otherwise be investigated, especially when the connectivity metric of the objective function is more emphasized and the LP relaxation does not offer a tight enough lower bound. Such restriction can also be beneficial by allowing the model to employ a much smaller big-M constant than previously possible, which can improve the LP relaxation and mitigate issues with numerical stability. 

When the $W+H$ constraint is introduced with a bound $P$, the big-M value can be set to $M = P + a_M$ without making otherwise feasible solutions infeasible. We set $a_M$ to the maximum of the minimum allowed distances between pairs of rectangles, $a_M = \max_{(i,j)} a_{i,j}$. This way, the constraints \eqref{pos1}-\eqref{rel2_eq} hold even in the most extreme cases. In experiments regarding the W+H constraint, we set the $P$ to half the perimeter of the previously found solution with additional slack to not restrict the solver too much. We discuss the obtained results in Section~\ref{red_perf}.

\section{Matheuristic as a Local Search}\label{matheuristic_section}
Given the initial solution, which can be provided either by the ILP solver with limited computation time or a suitable heuristic, we try to improve it using the variable-fixing MH. We refer to this improvement phase as intensification.
\subsection{Intensification}\label{intens}
\subsubsection{Rectangle Selection}
The choice of which variables should be fixed and which should remain flexible during intensification is crucial. Inspired by the job-window approach of \cite{Croce2014mh_varfix_sched}, we select a local group of rectangles $\mathcal{G}$. Given a position $(x,y)$ within the placement and size of the group $g$, the set $\mathcal{G}$ consists of $g$ rectangles closest to the point $(x,y)$. We define the 'proximity' metric of rectangle $i$ to point $(x,y)$ as:
\begin{equation}
    \mathrm{proximity}(x,y,i) = \max \left\{ |x_i-x|, |x_i + w_i - x|, |y_i-y|, |y_i+h_i-y|\right\}
\end{equation}

This way, selected rectangles should be located spatially close to each other, and when removed from the placement, mostly unfragmented empty space should appear. This should enable the solver to locally improve the connectivity by modifying the spatially local part of the placement. However, the positions and variants of the selected rectangles are not constrained, giving the solver the freedom to move them significantly if necessary. 

\subsubsection{ILP Intensification}
After the rectangle selection, the solver tries to improve the solution. The used ILP model corresponds to the one shown in Section~\ref{baseline_model}, so the feasibility of the solution is ensured. We fix the positions and variants of each rectangle $i \notin \mathcal{G}$; thus, their respective relative position variables $r_{i,j}^k$ or variant variables $s_i^k$ are not necessary. The selected rectangles belonging to $\mathcal{G}$ still have all their associated variables free. Therefore, the number of binary variables associated with $n$ rectangles decreases from:
\begin{equation}
    \sum_{i=1}^n m_i + 2 \cdot n \cdot (n-1)
\end{equation} to significantly smaller:
\begin{equation}
    \sum_{i \in \mathcal{G}} m_i + 2 \cdot g \cdot (g-1) + 4 \cdot g \cdot (n-g)
\end{equation}

Before the optimization, the solver is warm-started with the current placement. For a sufficiently small value of $g$, the solver is able to solve the restricted model optimally or at least find an improvement in a short time. Since the growing number of rectangles $n$ may slower intensification significantly, we impose a time limit on optimization.

\subsubsection{LP Fine Optimization}
To account for gaps between rectangles that can emerge by the variable fixing approach, we follow the previous step with LP optimization, which can lead to a lower value of HPWL and make the placement more compact. For each pair of rectangles, we find the least violated relative position constraint \eqref{pos1}-\eqref{rel2_eq}, and its associated variable $r_{i,j}^k$. Then, we optimize the original model of Section~\ref{baseline_model}, fixing the chosen relative position variables to~1 and the variant variables to select the variants present in the current solution. Thus, the model does not contain binary variables, and the optimization is done quickly, even for large instances. 

\subsubsection{Overall Intensification}
After each successful intensification iteration, the improved solution replaces the previous one. Then, a new selection point $(x,y)$ is sampled, and the process repeats until the computation budget is exhausted. In this paper, we generate the selection points by sampling uniformly from interval $\langle0; W\rangle$, $\langle0; H\rangle$ respectively. While such a simplistic strategy performed well, a more informed approach could yield better results.

\begin{figure}
	\centering
\begin{subfigure}{.49\textwidth}
  \centering
  \includegraphics[width=.98\linewidth]{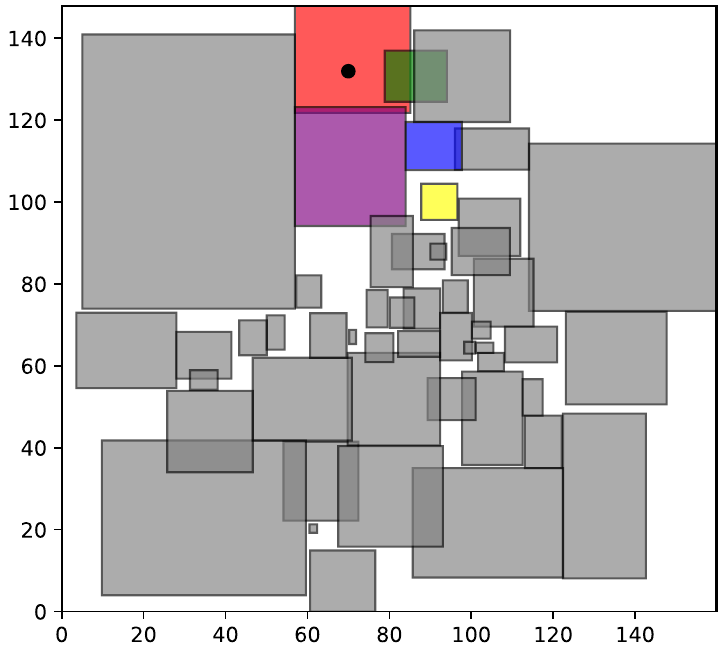}
  \caption{$\mathcal{L}_A = 307.9$, $\mathcal{L}_C = 3232.0$.}
\label{fig:i1}
\end{subfigure}
\begin{subfigure}{.49\textwidth}
  \centering
  \includegraphics[width=.98\linewidth]{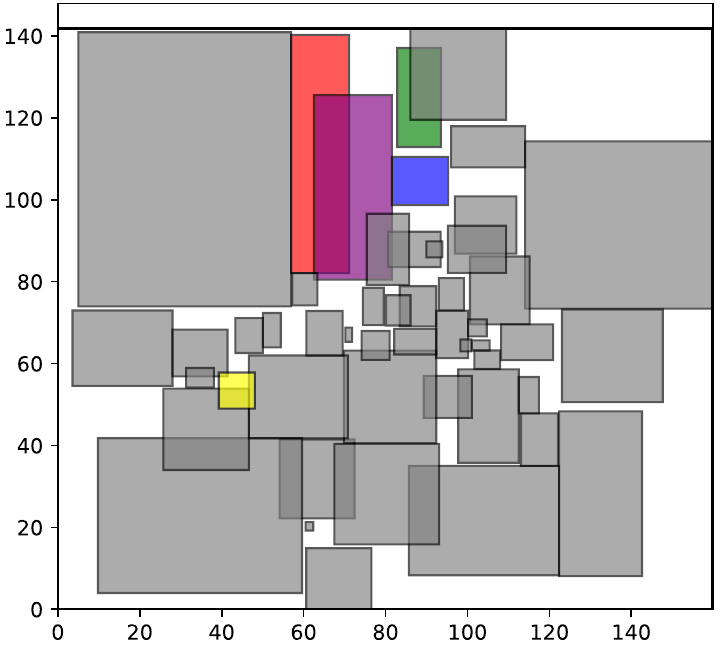}
  \caption{$\mathcal{L}_A = 301.8$, $\mathcal{L}_C = 3208.3$.}
   \label{fig:i2}
\end{subfigure}
    \caption{Placement before and after ILP intensification. The black dot shows where the selection position $(x,y)$ was sampled. Rectangles modified during the process are highlighted. See the decrease in height after intensification.}
   \label{fig_intens}
\end{figure}

The process of ILP intensification is demonstrated in Figs.~\ref{fig_intens}. The current placement is shown in Fig.~\ref{fig:i1}. The sampled position $(x,y)$, shown as a black dot, is located near the top side of the bounding box, and 5 rectangles were selected (red, purple, green, blue, and yellow). After the ILP intensification step, the new, improved placement is shown in Fig.~\ref{fig:i2}. We can see that the selected rectangles both moved and changed their variants. Both the half perimeter of the bounding box and the HPWL were decreased by this step, as is reported in the captions.

\subsection{Diversification}\label{divers}
While the ILP solver guarantees us that the local neighborhood of the current solution is thoroughly searched, the algorithm can get stuck in the local minimum. In that case, it is beneficial to divert from the current solution significantly and try to reach another potentially better local minimum. 

To perform a diversification step, we try to swap the positions of the rectangles so the overall placement changes, but we still try to keep the placement competitive. To do this, we create a swapping ILP model. In this model, each rectangle $i$ is associated with its centroid coordinates $(x^c_i,y_i^c)$, as well as its area $A_i = w_i \cdot h_i$. Note that chosen variant and coordinates of the rectangles are retrieved from the current solution. Then, the ILP model is formed as follows:

\begin{align}
     &\min ~~~  c_A \cdot \mathcal{L}_A + \frac{c_C}{\sum_{\forall e}c_e}\cdot\mathcal{L}_\mathrm{C} +  c_\xi \cdot \xi & \\
	   & \sum_{j=1}^n p_i^j = 1, ~~\sum_{j \in T_i} p_i^j = 1& \forall i \in \mathcal{I} \\
	   &  \sum_{i=1}^n p_i^j = 1& \forall j \in \mathcal{I} \\
& x_i^s = \sum_{j = 1}^n x_j^c \cdot p_i^j,~~~y_i^s = \sum_{j =1}^n y_j^c \cdot p_i^j & \forall i \in \mathcal{I}\\ 
     & \xi \ge N - (n - \sum_{i=1}^n p_i^i) \\
     & \xi \ge 0 \\
     &x_i^s,y_i^s \ge 0 & \forall i \in \mathcal{I} \\
     & p_i^j \in \left\{0,1\right\} & \forall i,j \in \mathcal{I}
\end{align}

Binary variable $p_i^j$ is equal to one if the rectangle $i$ should be placed to the position of the rectangle $j$ (thus, $p_i^i = 1$ means the rectangle $i$ does not move). To disallow the situation when a large rectangle would be placed in a position of the small one, we create a set of allowed swapping indices $T_i$ for each rectangle $i$. Note that $i \in T_i$ for each rectangle $i$.

\begin{equation}
    T_i =  \left\{j \in \mathcal{I} ~~\middle|~~ \frac{|A_i - A_j|}{\min\left\{A_i,A_j\right\}} \le A_\mathrm{diff} \right\}
\end{equation}

Maximum relative difference ($A_\mathrm{diff}=0.25$) limits the search space of the model significantly. Variables $x_i^s,y_i^s$ track the new centroid positions of the swapped rectangles that are used to calculate the half perimeter and connectivity criteria, using additional constraints shown in Section~\ref{baseline_model}. Finally, $\xi$ is used to penalize the insufficient number of swaps performed, i.e., when $p_i^i=1$ for too many rectangles. If the less than the expected minimum number of swaps $N$ is performed (we use $N=n/3$), additional penalty $c_\xi\cdot\xi$ is applied; we set the cost $c_\xi$ to quite a large value $\max\left\{W,H\right\}$, so the solver is motivated to perform the swaps.
 
After determining which swaps should be performed, we modify the current solution so the centroids of the swapped rectangles are moved to their associated positions. However, this can make the solution infeasible due to possible overlaps. To make the solution feasible, we use the original ILP model of Section~\ref{baseline_model} again. As in LP fine optimization of the intensification phase, we find the least violated relative position constraint for each pair of rectangles, and we warm start the solver with the corresponding variables set to 1. The values of variant variables are also obtained from the previous solution. The feasible result of the diversification phase is obtained by solving the model for a limited time. Afterward, we continue with intensification.

Since our intensification implementation does not exhaustively search all possible neighborhoods, we need a mechanism to decide when to perform the diversification and when to keep searching locally. Whenever the local search does not improve the solution's quality, we increment the counter. When the counter reaches 10, we perform the diversification and reset the counter. The counter is also reset when the improvement is achieved during the intensification.

\section{Experiments}\label{experiments}
\subsection{Methodology and Data}
We utilized the Gurobi ILP solver v9.5.1, using four threads in each experiment. The project was implemented using Python 3.7. Experiments were performed on an Intel Xeon E5-2690.

We generated several sets of instances inspired by the structure of real-life ones. Sets $S_{50}$ and $S_{100}$ were already discussed in our previous work \cite{icores23}. Additional sets $S_{200}$ and $S_{200}^\mathrm{sym}$ contain a larger number of rectangles, and the latter also contains several symmetry groups as a part of each instance. Each instance contains both the smaller rectangles, which only allow rotation, and larger ones with multiple variants. In total, 120 instances were evaluated. The computation time was fixed for each instance, depending on its set (shown in Table~\ref{tab:synth_overview}). When the MH was used, the initial solution was obtained by optimizing the original ILP model for a third of the computation time, and the rest was reserved for MH. The time required for warm starting the original model with the FDGD method was included in the total computation time. The costs in the criterion function were set to $c_A = 1.0$, and $c_C \in \left\{0.1,1.0,8.0\right\}$ respectively.

As baseline results, the methods proposed in \cite{icores23} were used. The baseline model without any improvement, denoted as \textbf{ILP}, was run only on the instance set $S_{50}$ and $S_{100}$, as it could not recover any solution for larger instances within the given runtime. FDGD warm-started variant \textbf{FDGD-ILP} solved all the instances.

\begin{table}
	\centering
	\caption{Description of synthetically generated instances.}
	\label{tab:synth_overview}
	\begin{tabular}{|c || c | c |c  |c|}\hline
	instance set & \# instances & \# rectangles & symmetry & comp. time\\
	\hline
	$S_{50}$ & 60 & 20, 30, 50 & No &  10 min\\
	$S_{100}$  & 20 & 100 & No &  20 min \\
	$S_{200}$  & 20 & 200 & No &  40 min \\
	$S_{200}^\mathrm{sym}$ & 20 & 200+ & Yes & 40 min \\\hline
	\end{tabular}
\end{table}

To compare the results obtained on the synthetically generated instances, we use the average relative difference (aRD) of the criterion, calculated for method $m$ and instance set $S$ as:
\begin{equation}
    \textrm{aRD}_S^m = \frac{1}{|S|}\cdot\sum_{i \in S}\frac{\mathcal{L}^{i, m} - \mathcal{L}^{i, {best}}}{\mathcal{L}^{i, {best}}} \cdot 100~[\%]
\end{equation}
where $\mathcal{L}^{i,m}$ is the value of criterion achieved on instance $i$ by method $m$, and $\mathcal{L}^{i,best}$ is the lowest value of criterion of among studied methods. Therefore, aRD refers to the ratio of the method's and best-known solution's criterion values averaged over the entire instance set. The best hits metric (BH) tells us how many times a specific method achieved the best-known value of the criterion.

\subsection{Performance with Redundant Constraints}\label{red_perf}
To study how the additional constraints affect the performance of the ILP solver, we performed experiments on instance sets $S_{50}$ and $S_{100}$. In the case of set $S_{100}$, only results for $c_C\in\left\{0.1,1.0\right\}$ are reported, as for $c_C=8.0$, not all methods found a feasible solution for each instance. The baseline \textbf{ILP} model is compared with symmetry-breaking one \textbf{SB-ILP} from Section~\ref{sb_ilp}, and the model \textbf{WH-ILP} using the W+H constraint from Section~\ref{wh_ilp}. Note that parameter $P$ used to define the W+H constraint was derived from the half perimeter of the feasible solution obtained using \textbf{FDGD-ILP}, which we increased by 20 \%. Furthermore, the studied instances did not contain symmetry groups; thus, utilizing symmetry breaking did not cause any problems. 

\begin{table}
	\centering
		\caption{Comparison of solutions obtained using baseline \textbf{ILP} model and the models with additional constraints, with reported values of aRD (BH) for each instance set and connectivity cost $c_C$.}
	\label{ilp_constrs}
		\begin{tabular}{|c || c| c| c || c| c|} 
        \hline
		   & \multicolumn{3}{c||}{$S_{50}$}& \multicolumn{2}{c|}{$S_{100}$}\\
          method & $c_C=0.1$ & $c_C = 1.0$ & $c_C = 8.0$  & $c_C=0.1$ & $c_C = 1.0$\\
		\hline
        \textbf{ILP} &  2.01 (19) & 1.69 (20) & 4.41 (22) & \textbf{1.81 (13)} & \textbf{2.29 (11)} \\
        \textbf{SB-ILP} & 1.90 (21) & 3.16 (21) & 5.55 (21) & 4.56 (12) & 6.07 (12) \\
        \textbf{WH-ILP} & \textbf{0.92 (31)} & \textbf{1.56 (33)} & \textbf{2.81 (29)} & - & - \\
        \hline
	\end{tabular}
 \end{table}

As shown in Table~\ref{ilp_constrs}, the results are rather inconclusive. The symmetry-breaking constraints help a little for $c_C=0.1$ on $S_{50}$ scenario, but lead to worse solutions on average. The W+H constraint leads to better solutions, but we were not able to find a feasible solution consistently for $S_{100}$ instances. In the case of the $c_C=0.1$ experiment on $S_{100}$ instance set, the feasible solution was found only for 10 of 20 instances. Furthermore, the average time needed to find the first feasible solution was 356 seconds. We concluded that imposing the upper bound on the half perimeter of the bounding box, and thus also on the big-M value, can improve the results. However, without passing the initial solution to a solver, the solver has a problem finding any feasible solution.

\subsection{Matheuristics on Synthetic Data}\label{mh_on_synth}
Our MH approaches rely on several parameters which may significantly influence the outcome of the local search. We fixed several parameters beforehand. When the diversification is used, we apply it after 10 non-improving intensification attempts. The maximum time reserved for each intensification and diversification optimization step was set to 10 seconds.

We performed experiments with four different MH settings. The settings \textbf{MH-5}, \textbf{MH-10}, and \textbf{MH-10D} used \textbf{FDGD-ILP} to find the initial solution for local search. Settings \textbf{MH-5} and \textbf{MH-10} relied only on intensification and differed in the number of rectangles $g$ selected to be optimized in each iteration (see Section~\ref{intens}). The first setting \textbf{MH-5} used $g=5$, and the second setting \textbf{MH-10} used $g=10$. The larger value of $g$ was not used, as the complexity of the larger model decreased the performance of the ILP solver significantly. The third setting \textbf{MH-10D} also used $g=10$ and employed diversification.

Finally, the remaining setting \textbf{MH-10B} used the baseline \textbf{ILP} method instead of the warm-started one to generate the initial solution. Then, it only uses intensification with $g=10$, thus being comparable to \textbf{MH-10}.


\subsubsection{Choice of suitable setting}
Firstly, we tried to determine how the value of $g$ and the use of diversification affects the results. We ran the experiments on all instance sets for all three values of the $c_C$. The experiments were performed with \textbf{MH-5}, \textbf{MH-10}, and \textbf{MH-10D} settings, and with \textbf{FDGD-ILP} serving as a baseline. The results are reported in Table~\ref{mh1}. 

We can see the baseline \textbf{FDGD-ILP} was outperformed on all instance sets. Furthermore, the improvement provided by MHs seems to be much more significant when the connectivity cost is high. This corresponds to the expected behavior of the intensification phase. Since we only free up to 10 rectangles in each iteration, and they are selected locally close to each other, there often remains a fixed rectangle that keeps the half perimeter of the bounding box unchanged. On the other hand, the connectivity of a single net can be significantly changed by moving even a single rectangle.

\begin{figure}
	\centering
\begin{subfigure}{.9\textwidth}
  \centering
  \includegraphics[width=.95\linewidth]{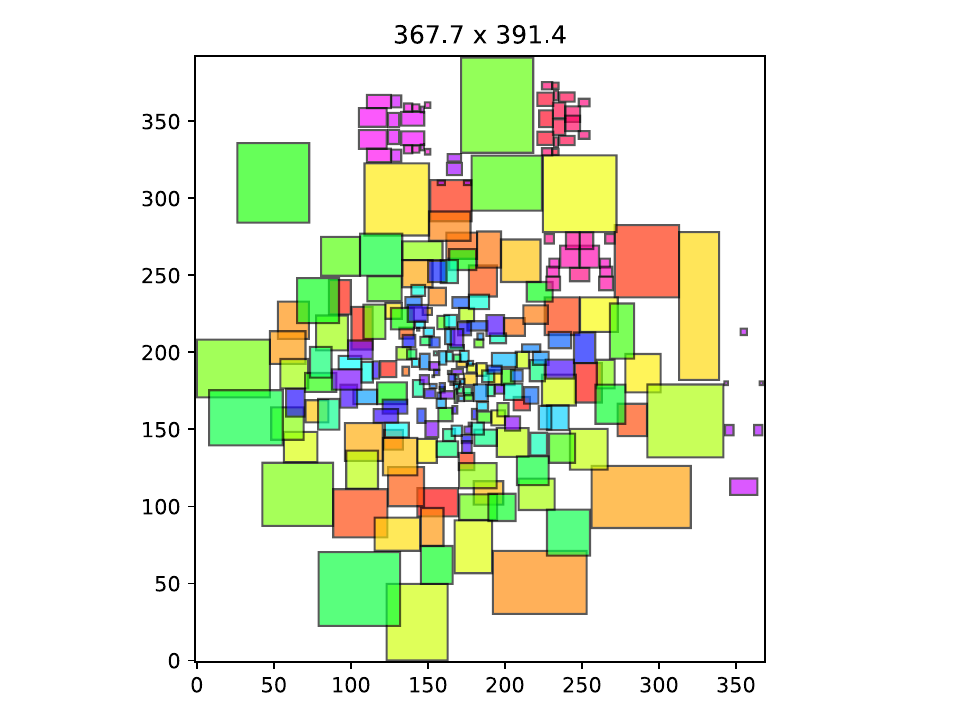}
  \caption{\textbf{FDGD-ILP} result, $\mathcal{L} = 1190.09$, area $=143894$, HPWL $=128885$.}
\label{fig:fdgdilp_vis}
\end{subfigure}
        \vskip\baselineskip
\begin{subfigure}{.9\textwidth}
  \centering
  \includegraphics[width=.95\linewidth]{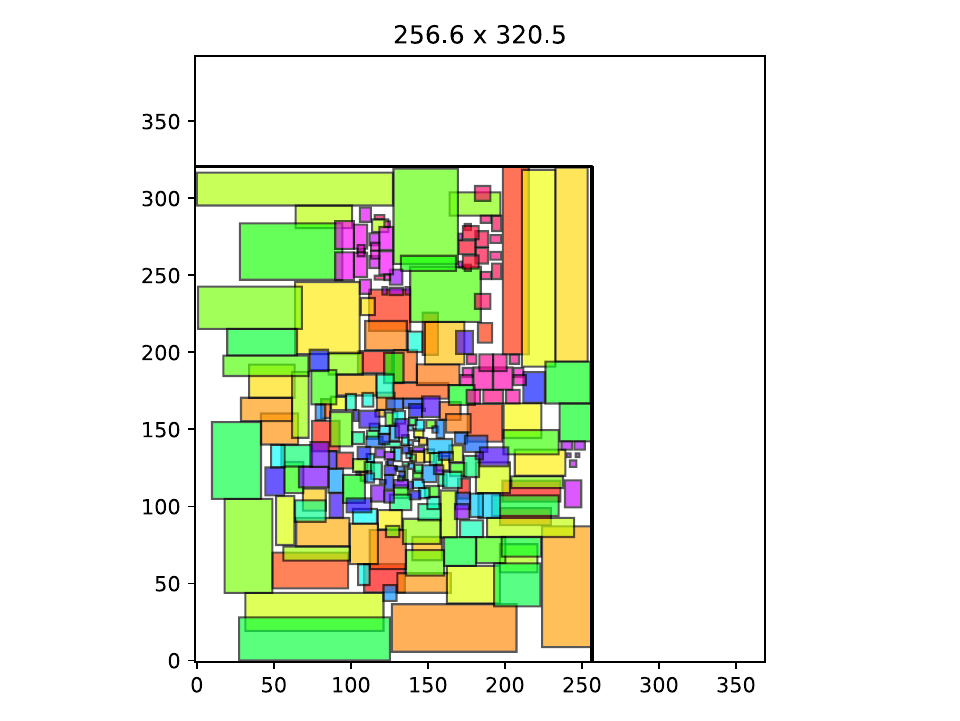}
  \caption{\textbf{MH-10} result, $\mathcal{L}=911.54$, area $=82235$, HPWL $=100000$.}
   \label{fig:mh10_vis}
\end{subfigure}
    \caption{Comparison of final placements obtained by \textbf{FDGD-ILP} and \textbf{MH-10} respectively, on instance from set $S_{200}^\mathrm{sym}$ with $c_C=1.0$. Both experiments' computation time was set to 2400 s.}
   \label{baseline_vs_mh}
\end{figure}

\begin{figure}
	\centering
    \includegraphics[width=0.97\linewidth]{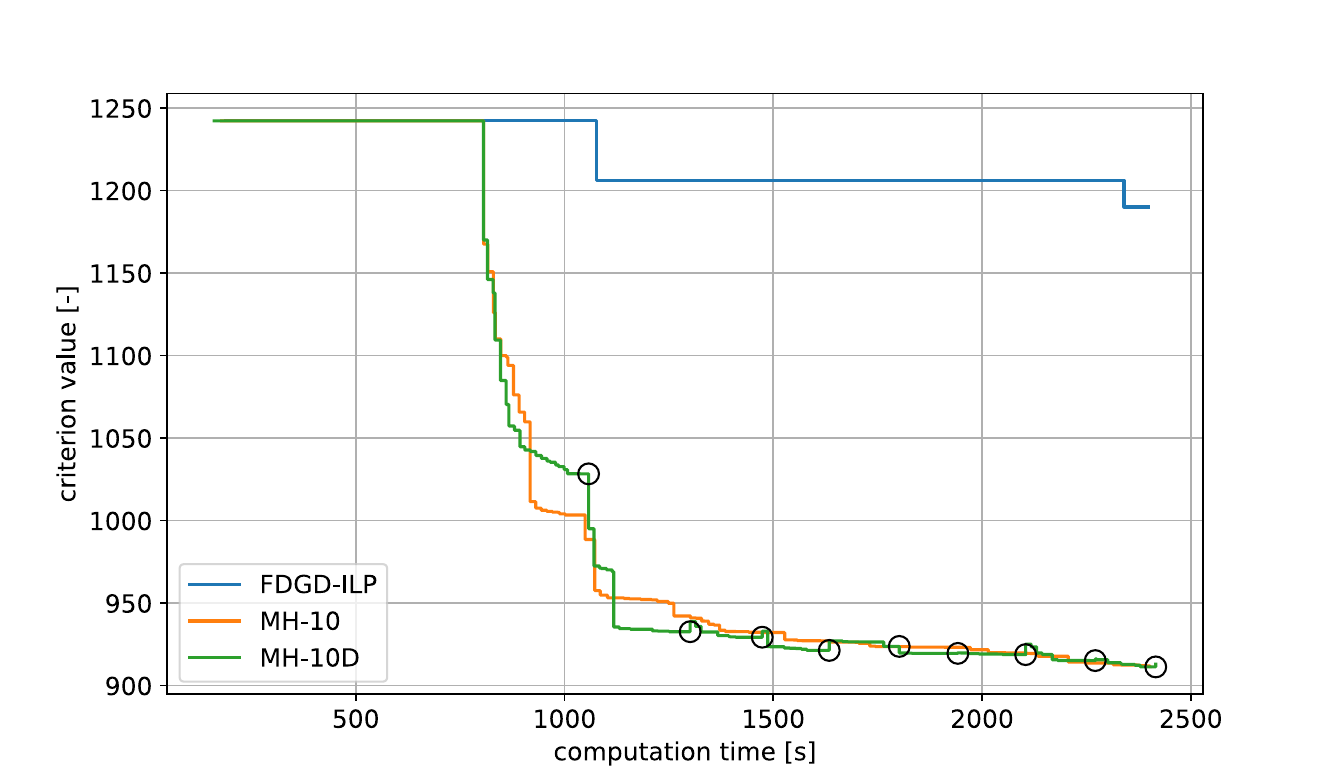}
    \caption{Value of criterion during optimization, for instance shown in Fig.~\ref{baseline_vs_mh}. Black circles show when \textbf{MH-10D} performed diversification.}
   \label{psym100graf}
\end{figure}

The improvements provided by the MHs are especially important in the case of instance set $S_{200}^\mathrm{sym}$, where the differences between the baseline results and the proposed methods are the largest - 30 \% on average. We believe that the main reason is the rigid handling of the symmetry groups our FDGD warm start uses. To create a feasible initial solution, each symmetry group is handled as a single entity, which, however, may lead to low-quality placement shown in Fig.~\ref{fig:fdgdilp_vis} (note, that we do not show internal devices inside the rectangles). Then, the solver cannot sufficiently improve the solution within the provided computation time due to the complexity of the model. On the other hand, the MH approach is able to decrease the value of the criterion significantly, and the overall placement looks more compact, see Fig.~\ref{fig:mh10_vis}. We also demonstrate this in Fig.~\ref{psym100graf}, which shows how the criterion value changes as the computation progresses. We can see that both shown MH settings, after their initialization phase, lower the criterion rapidly, while the solver optimizing the entire model struggles. This holds true even from the area-wise point of view; MH transforms the FDGD-produced circular placement to a more compact rectangular one.

\begin{table}
	\centering
		\caption{Comparison of different MH settings and \textbf{FDGD-ILP} baseline, with reported values of aRD (BH) for each instance set and connectivity cost $c_C$.}
	\label{mh1}
		\begin{tabular}{|c || c| c| c || c| c| c|} 
        \hline
		   & \multicolumn{3}{c||}{$S_{50}$}& \multicolumn{3}{c|}{$S_{100}$}\\
          method & $c_C=0.1$ & $c_C = 1.0$ & $c_C = 8.0$  & $c_C=0.1$ & $c_C = 1.0$ & $c_C = 8.0$ \\
		\hline
        \textbf{FDGD-ILP} & 2.06 (5)&  3.51 (0)  & 10.35 (1) & 2.08 (0) &  5.50 (0) & 12.51 (0) \\
        \textbf{MH-5} & 1.71 (6) & 2.34 (4) & 5.77 (2) & 0.85 (9) & 1.17 (7) & 1.81 (6) \\
        \textbf{MH-10} &\textbf{0.14 (44)} & \textbf{0.26 (43)} & 2.48 (15) &\textbf{0.48 (11)} &  \textbf{0.20 (13)} & \textbf{0.26 (14) }\\
        \textbf{MH-10D} & 2.12 (5) & 1.15 (13) & \textbf{0.70 (42)} & 3.80 (0) &  3.81 (0) & 3.82 (0) \\
        \hline
	\end{tabular}
 \begin{tabular}{|c || c| c| c || c| c| c|} 
        \hline
		   & \multicolumn{3}{c||}{$S_{200}$}& \multicolumn{3}{c|}{$S_{200}^\mathrm{sym}$}\\
          method & $c_C=0.1$ & $c_C = 1.0$ & $c_C = 8.0$  & $c_C=0.1$ & $c_C = 1.0$ & $c_C = 8.0$ \\
		\hline
        \textbf{FDGD-ILP} & 3.79 (3) & 8.02 (1) & 15.50 (0) & 27.10 (0) & 28.72 (0) & 31.61 (0) \\
        \textbf{MH-10} & \textbf{0.33 (16)} & \textbf{0.03 (19)} & \textbf{0.68 (19)} & \textbf{0.68 (15)} & \textbf{0.75 (11)} & 2.53 (1) \\
        \textbf{MH-10D} & 5.38 (1) & 5.83 (0) & 5.38 (1) & 1.51 (5) & 0.85 (9) & \textbf{0.04 (19)} \\
        \hline
        \end{tabular}
\end{table}

From the experiments on sets with less complex instances $S_{50},S_{100}$, we found out that while freeing only 5 rectangles leads to significant improvements and shorter optimization time per iteration, using $g=10$ yields better results on average. Therefore, we omitted the \textbf{MH-5} from the experiments on larger instances. Then we studied the effect of diversification. \textbf{MH-10} without diversification worked well in all cases, while the \textbf{MH-10D} was less predictable. However, for two instance sets with $c_C=8.0$, the \textbf{MH-10D} offered the best results, as is shown in Table~\ref{baseline_vs_mh}. Also, the larger diversification step can lead to significant improvements, as is illustrated in Fig.~\ref{psym100graf}, where the first time the diversification step is used (computation time 1100), the criterion drops significantly. We concluded that diversification offers advantages that could be more thoroughly exploited. However, due to the consistency of its results, we used the \textbf{MH-10} setting in the rest of the paper instead.

\subsubsection{Importance of the FDGD warm start}
After the previous experiments, we wanted to study whether it is still necessary to use the FDGD warm start to find the initial solution for MH. To do so, evaluated the original ILP model without warm start \textbf{ILP}, as well as its MH variant \textbf{MH-10B} on instances from $S_{50}$ and $S_{100}$. The results are reported in Table~\ref{mh_nofdgd}. 

\begin{table}
	\centering
		\caption{Comparison of FDGD-warm started and non-warm started MHs and ILP baselines, with reported values of aRD (BH) for each instance set and connectivity cost $c_C$.}
	\label{mh_nofdgd}
		\begin{tabular}{|c || c| c| c || c| c| c|} 
        \hline
		   & \multicolumn{3}{c||}{$S_{50}$}& \multicolumn{3}{c|}{$S_{100}$}\\
          method & $c_C=0.1$ & $c_C = 1.0$ & $c_C = 8.0$  & $c_C=0.1$ & $c_C = 1.0$ & $c_C = 8.0$ \\
		\hline
        \textbf{ILP} & 5.62 (2) & 10.29 (1) & 19.49 (1) & 19.76 (0) & 45.16 (0) & 46.98 (0) \\
        \textbf{FDGD-ILP} & 2.74 (1) & 3.78 (0) & 9.41 (1) & 2.21 (2) & 5.36 (0) & 12.33 (0)\\
        \textbf{MH-10B} & 1.93 (26) & 2.43 (15) & 2.95 (24) & 7.46 (4) & 5.60 (4) & 7.08 (2) \\
        \textbf{MH-10} & \textbf{0.82 (31)} & \textbf{0.52 (44)} & \textbf{1.62 (34)} &  \textbf{0.62 (14)} & \textbf{0.07 (16)}& \textbf{0.10 (18)}\\
        \hline
	\end{tabular}
 \end{table}

 From the provided table, we can see that the MH local search significantly improves the ILP baseline; it is even able to outperform the \textbf{FDGD-ILP} setting. When we compare the \textbf{MH-10B} with our main setting \textbf{MH-10}, we see that the FDGD warm start still provides some benefits. The warm-started variant of MH outperforms its non-warm-started counterpart, and this becomes more prevalent for more complex instances (where the \textbf{ILP} may not even find any solution). 

\subsection{Improvement on Real Life Instances}

Afterward, we studied how well MH works on real-life instances that were provided by industry partner STMicroelectronics and which we used previously in \cite{icores23}. 17 instances were provided, each consisting of up to 60 independent rectangles, and we ran two different experimental settings for each instance, either allowing or forbidding the use of pocket merging. Thus, the total number of experiments was 34. As in our previous work, the optimization was limited to 8 minutes. Three connectivity costs $c_C \in \left\{0.1, 1.0, 8.0\right\}$ were considered for each experiment. 

In Table~\ref{tab:realressize}, we report the metrics of manual designs and our solutions (the use of pocket merging depended on the manual design). The shown metrics are the half perimeter of the bounding box W+H, the placement area, and the connectivity metric HPWL. We found a solution dominating the metrics of the manual one for 12 out of 17 instances, matching our previous results. However, when we focus on the average ratios between automated and manual designs and compare them to results generated by \textbf{FDGD-ILP} in \cite{icores23}, we can see that we were able to quite significantly lower the connectivity while keeping the area and half-perimeter competitive. 

To highlight the differences between solutions found by \textbf{FDGD-ILP} and \textbf{MH-10}, we show Table~\ref{tab:comparison_real}. We can see that with the exception of the $c_C=0.1$ scenario, the MH approach, on average, reduced the criterion of the final solution and found the better solution in a majority of the cases. This again corresponds to the observations we made in Section~\ref{mh_on_synth}. The real-life instances also contain only up to 60 rectangles, and as we have shown, the effect of the MH shows off when the instances are more complex.

Furthermore, note the values reported in columns DOM. These correspond to a number of occurrences when the method found a solution that had both a smaller area and HPWL than the solution found by the other method; such a solution is objectively better given our two main metrics. We can see that \textbf{MH-10} was able to do so in more cases, which again highlights the power of local search performed by the ILP solver.

\begin{figure}
	\centering
\begin{subfigure}{.9\textwidth}
  \centering
  \includegraphics[width=.53\linewidth]{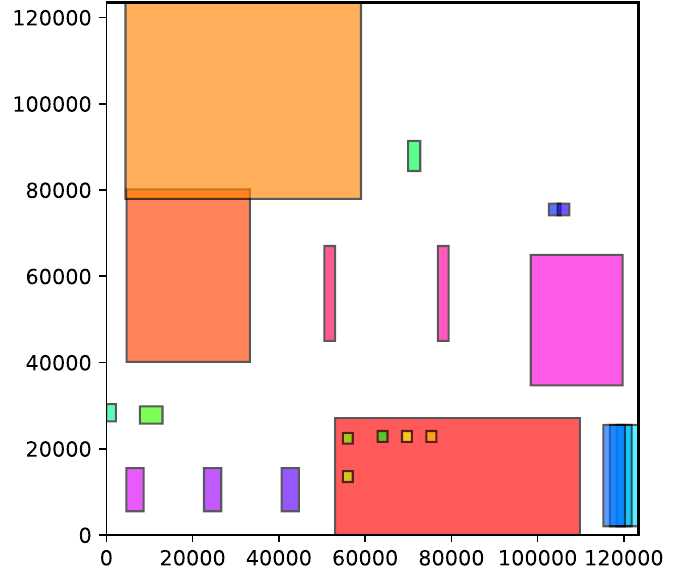}
  \caption{Manual design, area $=15235 \mu\textrm{m}^2$, HPWL $=2399 \mu$m.}
\label{fig:rl_man}
\end{subfigure}
        \vskip\baselineskip
\begin{subfigure}{.9\textwidth}
  \centering
  \includegraphics[width=.53\linewidth]{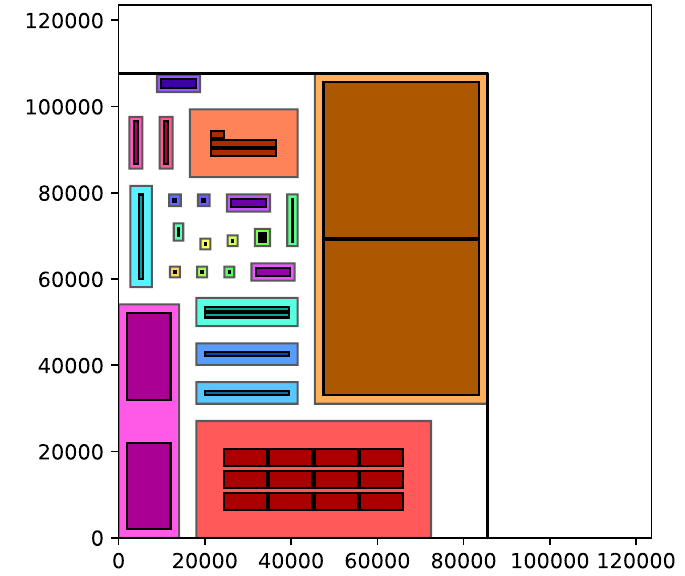}
  \caption{\textbf{FDGD-ILP}, area $=9212 \mu\textrm{m}^2$, HPWL $=1898 \mu$m.}
   \label{fig:rl_ilp}
\end{subfigure}       
\vskip\baselineskip
\begin{subfigure}{.9\textwidth}
  \centering
  \includegraphics[width=.53\linewidth]{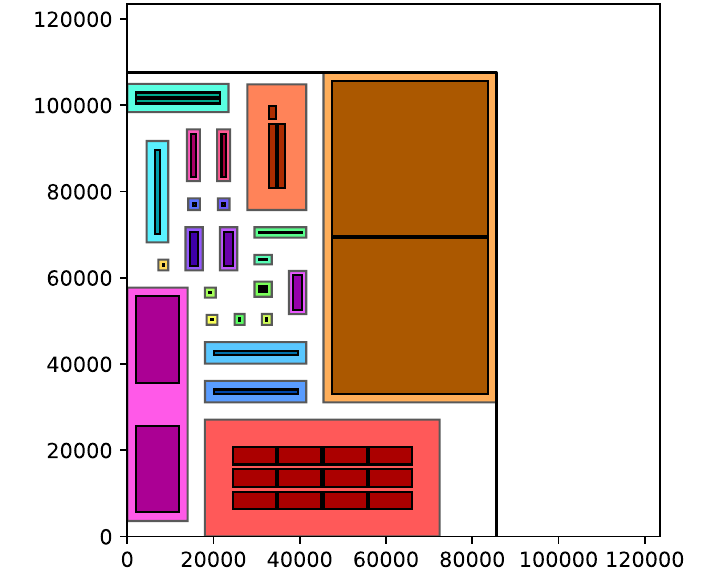}
  \caption{\textbf{MH-10}, area $=9212 \mu\textrm{m}^2$, HPWL $=1720 \mu$m.}
   \label{fig:rl_mh}
\end{subfigure}
\caption{Comparison of manual and automated placements, obtained for $c_C = 0.1$. Shown instance corresponds to the 14th row in Table~\ref{tab:realressize}.}
   \label{example_rl}
\end{figure}

We illustrate the mentioned results with Figs.\ref{example_rl}, which show instance number 14. Three figures correspond to the manual solution, the solution obtained using \textbf{FDGD-ILP}, and finally using \textbf{MH-10}. Note that the manual design does not show the positions of the physical devices within the rectangles. We can see that both automatically generated solutions dominated the manual design. The differences between both automated solutions are subtle; their areas are actually equal. However, even these subtle changes in the positions of smaller rectangles are enough to dramatically decrease the HPWL in the case of the solution generated by \textbf{MH-10}.

\begin{table} 
\centering
\caption{Values of $W+H$ in $\mu$m, area in $\mu\textrm{m}^2$ and HPWL in $\mu$m for each instance, and average ratios of automated and manual metrics, obtained using \textbf{MH-10}. The average ratios obtained by \textbf{FDGD-ILP} in \cite{icores23} are shown in the last row for comparison. Solutions dominating the manual one, given all three metrics, are highlighted.}
\label{tab:realressize}
    \begin{adjustbox}{width=0.95\textwidth}
\begin{tabular}{|c||c c c||c c c|c c c|c c c|}\hline
	 & \multicolumn{3}{c||}{manual} & \multicolumn{9}{c|}{\textbf{MH-10}} \\
	& \multicolumn{3}{c||}{}& \multicolumn{3}{c|}{$c_{conn} = 0.1$} & \multicolumn{3}{c|}{$c_{conn} = 1.0$} & \multicolumn{3}{c|}{$c_{conn} = 8.0$}  \\
	instance &  W+H & area & HPWL &  W+H &  area & HPWL & W+H &  area & HPWL &   W+H &  area & HPWL\\
	\hline
1& 158& 6118& 1850& 157& 6172& 1636& 157& 6183& 1562& 166& 6889& 1478\\
2& 116& 2710& 1784& \textbf{88}& \textbf{1936}& \textbf{1024}& \textbf{91}& \textbf{2070}& \textbf{928}& 106& 2757& 797\\
3& 106& 2650& 906& \textbf{85}& \textbf{1779}& \textbf{660}& \textbf{89}& \textbf{1968}& \textbf{654}& \textbf{92}& \textbf{2119}& \textbf{547}\\
4& 129& 4096& 812& \textbf{112}& \textbf{3117}& \textbf{782}& \textbf{114}& \textbf{3256}& \textbf{717}& 131& 4064& 662\\
5& 207& 8972& 13797& \textbf{159}& \textbf{6351}& \textbf{9955}& \textbf{165}& \textbf{6789}& \textbf{8141}& \textbf{169}& \textbf{7120}& \textbf{7863}\\
6& 178& 7698& 4039& \textbf{169}& \textbf{7167}& \textbf{3666}& \textbf{167}& \textbf{7009}& \textbf{3647}& \textbf{174}& \textbf{7224}& \textbf{3615}\\
7& 168& 6580& 2908& 164& 6756& 2633& 168& 7093& 2314& 173& 7466& 2307\\
8& 173& 7294& 1501& \textbf{160}& \textbf{6399}& \textbf{1224}& \textbf{169}& \textbf{6973}& \textbf{1068}& \textbf{173}& \textbf{7139}& \textbf{1093}\\
9& 243& 14129& 4705& \textbf{225}& \textbf{12647}& \textbf{4205}& \textbf{234}& \textbf{13664}& \textbf{4003}& 241& 14487& 3882\\
10& 205& 10214& 28386& 191& 9093& 38626& 194& 9446& 32363& 236& 13714& 24930\\
11& 225& 9922& 28527& 197& 9356& 29074& 205& 10313& 17864& 241& 13717& 13210\\
12& 155& 5953& 3824& \textbf{123}& \textbf{3803}& \textbf{2315}& \textbf{126}& \textbf{3937}& \textbf{2162}& 159& 6298& 1597\\
13& 162& 6511& 2061& \textbf{153}& \textbf{5855}& \textbf{1822}& \textbf{155}& \textbf{6002}& \textbf{1665}& \textbf{155}& \textbf{6008}& \textbf{1693}\\
14& 247& 15235& 2399& \textbf{193}& \textbf{9212}& \textbf{1720}& \textbf{193}& \textbf{9263}& \textbf{1557}& \textbf{211}& \textbf{10657}& \textbf{1363}\\
15& 123& 3758& 1619& 115& 3309& 1817& 113& 3178& 1852& 116& 3385& 1712\\
16& 232& 12397& 2676& \textbf{215}& \textbf{11551}& \textbf{1973}& \textbf{223}& \textbf{12318}& \textbf{1792}& \textbf{221}& \textbf{12143}& \textbf{1944}\\
17& 247& 12525& 4586& \textbf{225}& \textbf{12172}& \textbf{3313}& 235& 13708& 3008& 252& 15790& 2964\\
\hline
\makecell{avg ratio \\ \textbf{MH-10}}
& 1.00 & 1.00 & 1.00 & 0.89 & 0.84 & 0.86 & 0.91 & 0.89 & 0.77 & 0.98 & 1.02 & 0.71 \\
\hline
\hline
\makecell{avg ratio \\ \textbf{FDGD-ILP} \cite{icores23}}& 1.00 & 1.00 & 1.00 & 0.88 & 0.84 & 0.93 & 0.91 & 0.89 & 0.82 & 0.99 & 1.04 & 0.74 \\\hline
\end{tabular}
\end{adjustbox}
\end{table}

\begin{table} 
\centering
\caption{Comparison of \textbf{FDGD-ILP} and \textbf{MH-10} for all 34 experiments performed on real-life instances. DOM shows in how many cases the method dominated the other one with respect to both the area and HPWL.}
\label{tab:comparison_real}
\begin{tabular}{|c||c |c|c| c|c |c|}
\hline
	 & \multicolumn{2}{c|}{$c_C = 0.1$} & \multicolumn{2}{c|}{$c_C=1.0$} & \multicolumn{2}{c|}{$c_C=8.0$} \\
	 & aRD (BH) & DOM & aRD (BH) & DOM & aRD (BH) & DOM\\
	\hline
 \textbf{FDGD-ILP} & \textbf{0.46 (22)} & 1 & 1.82 (11) & 1 & 3.84 (14) & 0 \\
 \textbf{MH-10} & 0.60 (12) & 8 & \textbf{0.11 (23)} & 10 &\textbf{ 0.49 (20)} & 11\\
 \hline
\end{tabular}
\end{table}

\section{Conclusion}\label{conclusion}

In this paper, we extended our previous work on the automation of the placement of AMS ICs. We studied the effect of additional redundant constraints on the performance of the state-of-the-art ILP solver. While the symmetry-breaking constraints did not enhance the solver's performance, imposing an additional constraint on the maximum value of half perimeter of the placement led to improvement on the smaller instances. However, for larger instances, such constraint made the solver unable to find any feasible solution in a given computation time, even though the bound was derived from a known feasible solution. Therefore, we need to provide a solver with an initial solution if we would like to exploit the half-perimeter constraint in the future.

Our experiments with MHs were more successful. We proposed applying the ILP solver to perform a local search in the created placement. The intensification phase of the MH relied on freeing variables associated with a few spatially close rectangles while fixing the other. We showed an additional ILP model that we used to perform the diversification step, to diverge further from the current solution when the local minimum is reached. We evaluated several different MH settings on synthetically generated instances. We concluded that using intensification only and freeing 10 rectangles in each iteration led to the best results overall. However, the potential benefits of diversification cannot be overlooked, but its application would probably require a more advanced control mechanism than presented in our paper. Ultimately, we significantly improved our previous results, obtained using FDGD-warm started ILP, on large instances with 200 and more rectangles, especially when symmetry groups are present in the instance.

Finally, we created automatically generated placements for real-life instances provided by our industry partner STMicroelectronics. We could compare our results with the manually created benchmarks and our previous results. We were again able to outperform both the area and the HPWL in the case of 12 instances; furthermore, we were able to reduce the average value of HPWL even further while keeping the area metric unaffected. When we analyzed the improvement against our previous results more closely, we found that the MH approach dominated its ILP-only counterpart regarding both the HPWL and area in one-third of the experiments performed on real-life instances. This again suggests that the use of MH could be beneficial not only in the specific domain of AMS IC placement but in the domains of packing and cutting as well, where only the area is minimized.

\subsubsection{Acknowledgements} 
This work was supported by the Grant Agency of the Czech Republic under the Project GACR 22-31670S. This work was co-funded by the European Union under the project ROBOPROX - Robotics and advanced industrial production (reg. no. CZ.02.01.01/00/22\_008/0004590). We would like to thank the STMicroelectronics company, namely Dalibor Barri and Patrik Vacula, for providing real-life instances and helpful discussions about a problem.

%
%
%
\bibliographystyle{splncs04}
\bibliography{bibliography}
\end{document}